\documentclass{article}

\usepackage{graphicx}

\def\CC{$^{12}$C/$^{13}$C~}
\begin{document}

\title{The Carbon Abundance and \CC Isotopic Ratio in the Atmosphere of 
Arcturus from 2.3 micron CO Bands}

\author{Ya. V. Pavlenko\thanks{Main Astronomical Observatory, National Academy of Sciences of Ukraine, 
27 Zabolotnogo, Kiev, 03680 Ukraine; http://www.mao.kiev.ua/staff/yp}}

\maketitle

\begin{abstract}

Absorption lines of the $^{12}$CO and $^{13}$CO molecular bands 
($\Delta v$ = 2) at  
2.29 -- 2.45 micron are modelled in spectrum of Arcturus (K2III). 
We compute a grid of model 
atmospheres and synthetic spectra for giant of
Teff = 4300, log g = 1.5, and the elemental abundances of Peterson et 
al. (1993), but abundances of carbon,
oxygen and the carbon isotopic ratio, \CC are 
varied in our computations. The computed spectra are fitted to the 
observed spectrum of Arcturus from the atlas of Hinkle et al. (1995). 
The best fit to observed spectrum is achieved for 
log N(C) = -3.78 $\pm$ 0.1, \CC = 8 $\pm$ 1. A dependence of 
the determined \CC vs. 
log N(C) and log N(O) in atmospheres of red giants is discussed.

\end{abstract}

\clearpage


\begin{figure}
\begin{center}
\includegraphics [width=8cm]{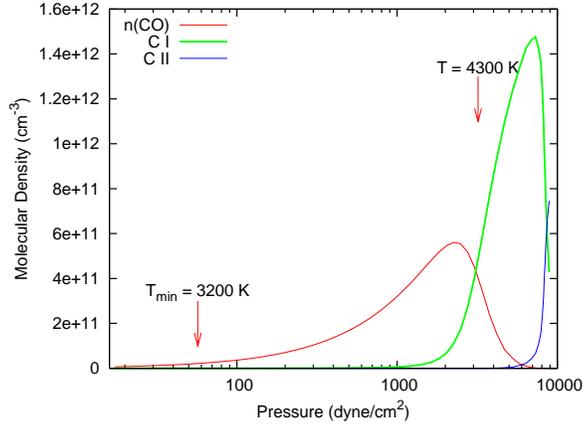}
\end{center}
\caption[]{\label{_nCO} 
 Molecular densities of CI, CII and CO in the atmosphere 
with Teff = 4300 K and log g = 1.5. The elemental abundances were taken 
from Peterson et al. (1993), abundance of carbon log N (C) = -3.78.
}
\end{figure}

\begin{figure}
\begin{center}
\includegraphics [width=8cm]{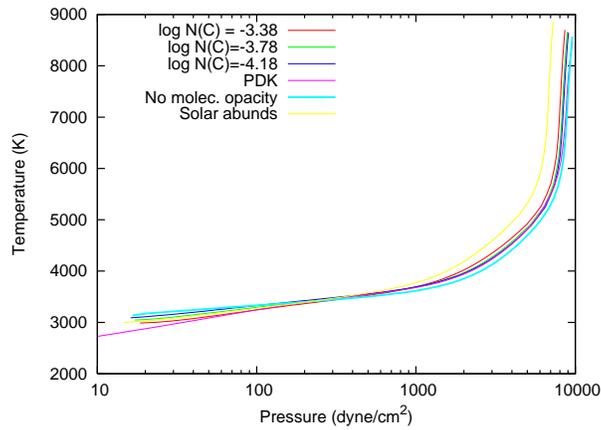}
\end{center}
\caption[]{\label{_modelc} 
Temperature distributions in our model atmospheres of Arcturus 
computed for i)
different carbon abundances, ii) solar chemical 
composition, iii) our model atmosphere computed taking into 
account only opacity in continuum and atomic lines, iv) the 
temperature distribution in the PDK model (Peterson at al. 1993).
}
\end{figure}

\clearpage

\begin{figure}
\begin{center}
\includegraphics [width=8cm]{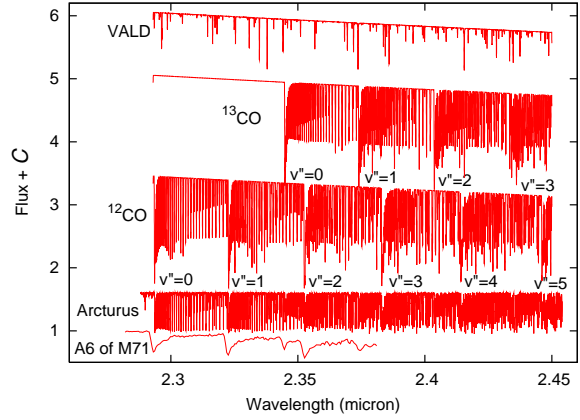}
\end{center}
\caption[]{\label{_ident} 
Identification of CO bands and atomic lines in the spectrum of 
Arcturus in the 
modelled wavelength range. For comparison, we show the observed spectra 
of Arcturus from Hinkle et al. (1995) and of the giant A6 in the 
globular cluster M71 (Pavlenko et al. 2003).
}
\end{figure}

\begin{figure}
\begin{center}
\includegraphics [width=8cm]{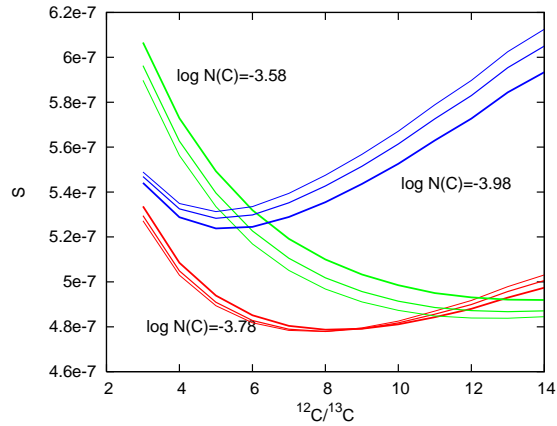}
\end{center}
\caption[]{\label{_OC} 
Minimum of $S=\Sigma (F^{obs}_i - F^{synt}_i)$, here $F^{obs}_i$ and 
$F^{synt}_i$ are observed and computed fluxes,  
(see Pavlenko et al. 2003 for more detailed explanation)
allows to determine  the best values of \CC and
carbon abundance. In all cases, the effective 
temperature and gravity are Teff = 4300 K and log g = 1.5. The abundances 
of other elements are from Peterson et al. (1993).
}
\end{figure}
                            
\clearpage

\begin{figure}
\begin{center}
\includegraphics [width=8cm]{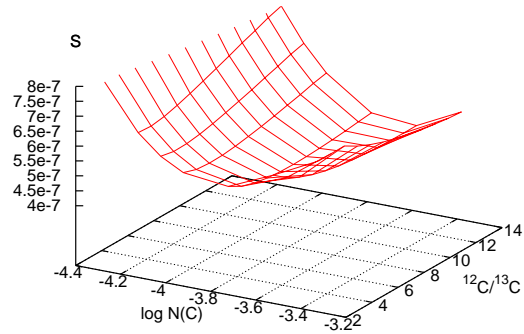}
\end{center}
\caption[]{\label{_3D} 
S values obtained from the fits of
synthetic spectra computed for a series of model atmospheres of 
Teff = 4300 K and log g = 1.5 and
different values of log N(C) and \CC
to the observed spectrum of Arcturus 
(Hinkle et al. 1995). The abundances of other 
elements were taken from Peterson et al.(1993).
}
\end{figure}

\begin{figure}
\begin{center}
\includegraphics [width=8cm]{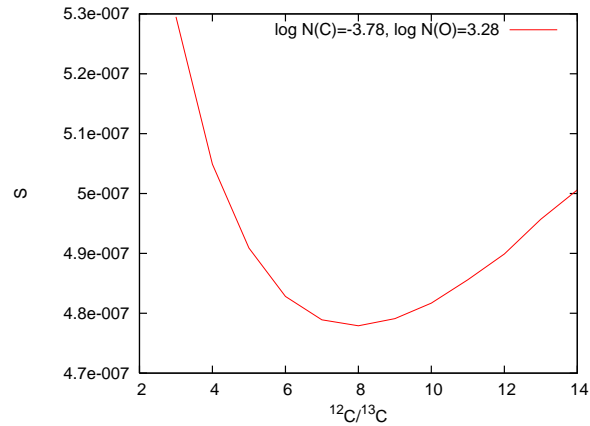}
\end{center}
\caption[]{\label{_bestfit} 
$S$ vs. \CC dependence provides the best-fit 
carbon isotope ratio ratio for Arcturus \CC = 8 +/- 1.
Abundance of carbon is log N(C) = -3.78.
}
\end{figure}

\clearpage

\begin{figure}
\begin{center}
\includegraphics [width=8cm]{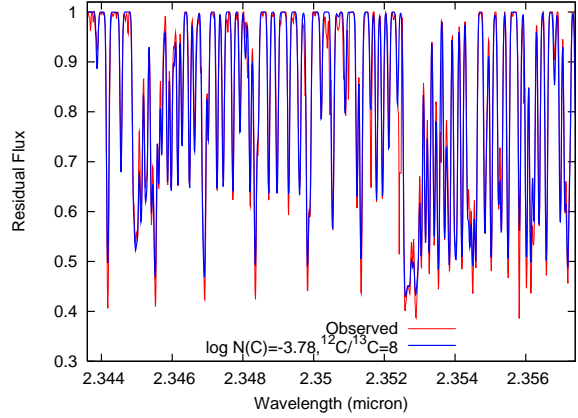}
\end{center}
\caption[]{\label{_fitto} 
Fit of the computed spectrum for our model atmosphere with 
log N(C) = -3.78, log N(O) = -3.21, \CC = 8 to the
observed spectrum of Arcturus ( Hinkle et al. 1995)
}
\end{figure}

\begin{figure}
\begin{center}
\includegraphics [width=8cm]{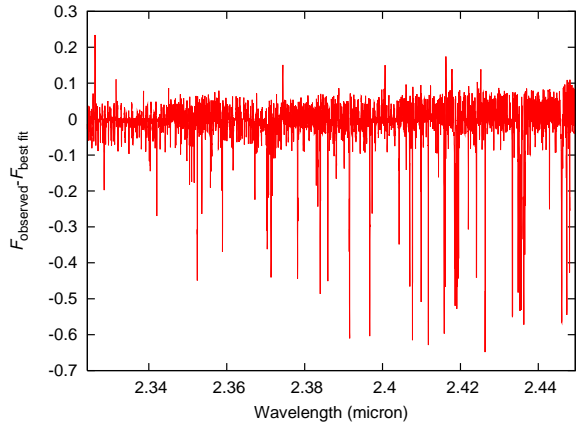}
\end{center}
\caption[]{\label{_diff} 
The difference between the observed and computed fluxes
of Arcturus.  Theoretical  spectrum was computed for our model 
atmosphere of Teff = 4300 K and log g = 1.5 with log 
N(C) = -3.78, log N(O) = -3.21, and \CC = 8. 
The other abundances  are from Peterson et al. (1993).
}
\end{figure}

\end{document}